\documentclass[prl,twocolumn,showpacs,groupedaddress,superscriptaddress,nofootinbib,floatfix,preprintnumbers,longbibliography]{revtex4-1}

\usepackage{amssymb,amsmath,graphicx,color}
\usepackage{bm}
\usepackage[tight]{subfigure}
\usepackage[export]{adjustbox}
\usepackage{braket}
\usepackage[colorlinks=true,citecolor=blue,linkcolor=red]{hyperref}
\usepackage{multirow}
\usepackage{rotating,booktabs}
\usepackage[verbose]{placeins}

\usepackage[normalem]{ulem}

\begin{document}
	
	\title{Phonon-induced pairing in quantum dot quantum simulator}
	
	\author{Utso Bhattarcharya }
	\affiliation{
		ICFO-Institut de Ciencies Fotoniques, The Barcelona Institute of
		Science and Technology, Castelldefels (Barcelona) 08860, Spain.}
	\affiliation{Max-Planck-Institut für Quantenoptik, D-85748 Garching, Germany}
	\author{Tobias Grass }
	\affiliation{
		ICFO-Institut de Ciencies Fotoniques, The Barcelona Institute of
		Science and Technology, Castelldefels (Barcelona) 08860, Spain.}
	\author{Adrian Bachtold }
	\affiliation{
		ICFO-Institut de Ciencies Fotoniques, The Barcelona Institute of
		Science and Technology, Castelldefels (Barcelona) 08860, Spain.}
	\author{Maciej Lewenstein }
	\affiliation{
		ICFO-Institut de Ciencies Fotoniques, The Barcelona Institute of
		Science and Technology, Castelldefels (Barcelona) 08860, Spain.}
	\affiliation{ICREA, Pg. Lluis Companys 23, 08010 Barcelona, Spain}
	\author{Fabio Pistolesi }
	\affiliation{Univ. Bordeaux, CNRS, LOMA, UMR 5798, F-33400 Talence, France }
	
	\begin{abstract}
		Quantum simulations can provide new insights into the physics of strongly correlated electronic systems.
		A well studied system, but still open in many regards, is the Hubbard-Holstein Hamiltonian, where electronic repulsion is in competition with attraction generated by the electron-phonon coupling.
		In this context we study the phase diagram of four quantum dots in a suspended carbon nanotube and coupled 
		to its flexural degrees of freedom. 
		The system is described by a Hamiltonian of the Hubbard-Holstein class, where electrons on different sites interact with the same phonon.
		We find that the system presents a transition from the Mott insulating state to a polaronic state, with the appearance of pairing correlations and the breaking of the translational symmetry.
		Our study shows that this system thus constitutes a relevant example of a correlated system that could be studied by experimental realization. 
	\end{abstract}
	
	\maketitle
	As for today, quantum simulators are the unique systems that can address, deepen our understanding, and ultimately solve with {\it quantum advantage} 
	challenging problems of contemporary science: from quantum many body dynamics, through static and transient high $T_c$ superconductivity to design of new materials. 
	As an important example one can think to the Hubbard model, a paradigm of strongly correlated systems, and that has been investigated 
	through a number of experimental platforms. 
	Such platforms include arrays of three and four dots in semiconductor heterostructures \cite{Hensgens2017,Dehollain2020}, as well as various 
	setups used in the ultracold-atom community \cite{Greiner,Greiner0,Greiner1,Greiner2,Greiner3,Trotzky2008,Gorg2018,Salomon2019,Nichols2019}. Recent efforts in ultracold-atom physics focus on one side on quantum simulators of exotic topological order and spin liquids (cf. Ref.~\onlinecite{topological}). On the other side, 
	dynamical lattices \cite{PhysRevLett.111.080501,PhysRevLett.121.090402} and lattice gauge theory models are in the center of interest, where additional dynamical degrees of freedom live on the bonds of the lattice (for reviews see \cite{Banyuls,Aidelsburger}). 
	
	An important goal for such quantum simulators is the study of electron-phonon class of models (EPCM), which play a crucial role for understanding strongly correlated condensed matter systems. These models exhibit, in addition to electron-electron (e-e) interactions, coupling between electrons and lattice degrees of freedom, in form of  quantized lattice displacements or phonons, either locally as in the Hubbard-Holstein model \cite{beni1974} or through long range electron-phonon (e-p) interactions as captured by the Hubbard-Froehlich model \cite{Alexandrov2002}. The e-p interactions can generate effective attractive e-e interactions, via the Cooper pairing channel in conventional superconductors \cite{bardeen57} and therefore, directly compete with the repulsive e-e interactions present. Accordingly, such systems can host a plethora of phases like superconductivity, antiferromagnetism, charge density waves, etc.\cite{karakuzu_superconductivity_2017}. The role of the competing interactions for the creation of high-temperature superconductivity and the pseudogap states in strongly correlated low-dimensional materials is an active topic of investigation. \cite{lanzara2001,julia2020,shen2004}. Moreover, the interplay between the e-p interaction and the Coulombic e-e repulsion is of great importance for unconventional superconductors such as alkali-metal-doped fullerides \cite{Takabayashi2009}, pnictides \cite{delaCruz2008,PhysRevLett.104.157001},  and aromatic superconductors \cite{Mitsuhashi2010} as well.
	
	Although many numerical techniques, such as quantum Monte Carlo \cite{clay2005}, the density-matrix renormalization group (DMRG) \cite{fehske2008}, variational ansatz \cite{alder1997}, the dynamical mean-field theory (DMFT) \cite{werner2007} and  density-matrix embedding theory (DMET) \cite{sandhoefer2016}, have been devised over the years for the EPCM, there is still a distinct lack of reliable numerical and analytical solutions that are not restricted to only a portion of its parameter space. 
	Therefore, a tunable quantum simulator which would allow for exploring these models without the need for making some approximations is highly sought after, and it might reveal the entire richness present in such models.
	
	%
	%
	Here, we discuss in detail, within practical experimental limits, a new approach of simulating an EPCM with e-e and e-p interactions exploiting electromechanical devices.
	Such systems have been employed with great success to couple mechanical modes to quantum electron transport.
	Several transport regimes have been studied, such as single-electron tunneling \cite{steele_strong_2009,lassagne_coupling_2009,Woodside1098,knobel_nanometre-scale_2003,naik_cooling_2006,ganzhorn_dynamics_2012,benyamini_real-space_2014, pirkkalainen_cavity_2015,ares_resonant_2016,khivrich_nanomechanical_2019,wen_coherent_2020,blien_quantum_2020,vigneau_ultrastrong_2021} Kondo \cite{gotz2018,urgell_cooling_2020} and the quantum Hall effect \cite{chen_modulation_2016}. 
	In these systems the electrons can be localized in one (or two) quantum dots  and they  interact electrostatically  with one or several mechanical modes. 
	The strong and controllable localization of the charge, with reduced screening, allows one to reach very large e-p coupling 
	constant. This leads to strong back-action on the oscillator with the predicted formation of polaronic states and suppression of the conductance 
	\cite{galperin_hysteresis_2005,koch_franck-condon_2005,pistolesi_self-consistent_2008,micchi_mechanical_2015} and observed softening of the mechanical resonating frequency \cite{lassagne_coupling_2009,steele_strong_2009}.
	These systems offer great opportunities for quantum simulations, since  several parameter of the 
	resulting Hamiltonian can be tuned independently, either at the fabrication stage, or during the experiment.  
	Advances in nanofabrication should soon enable the fabrication of several quantum dots (more than four) in a suspended carbon nanotube. 
	By tuning the nearby gate voltages, one can tune by a large amount the hopping term between the quantum dots, the local potential, and the e-p coupling. 
	The system can thus be used to simulate Hamiltonians that fall in the EPCM.
	Even if the system is of finite size, in comparison 
	with the bulk counterparts of the same model, 
	one has the advantage that local physical quantities can be tuned and quantified, leading to the possibility to gain a totally new insight on the physical problem. 

\begin{figure}[t]
		\begin{center}
			\includegraphics[width=8cm]{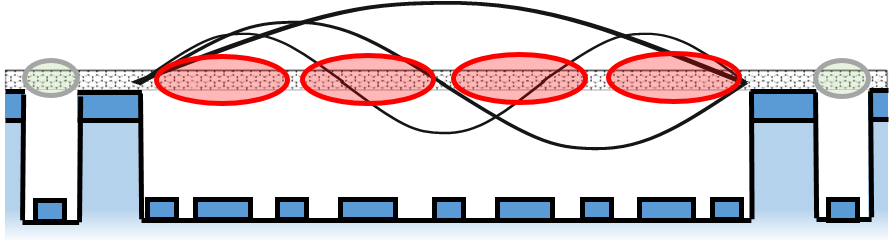}
		\end{center}
		\caption{Schematic of the proposed setup. Four quantum dots in red are electrostatically defined along a suspended nanotube using the gate electrodes at the bottom of the trench. The electron states of the four quantum dots are coupled to the nanotube mechanical eigenmodes depicted as black lines. Real-time charge sensing of the quantum dot array can be experimentally monitored with the sensing dots in grey defined on the sides \cite{Hensgens2017}.  
		}
		\label{fig1}
	\end{figure}

	In this Letter we study the behavior of a system consisting of four quantum dots coupled to a set of phonons as a function 
	of the e-p coupling constant and the hopping integral. 
	For vanishing hopping, the problem can be exactly solved. Here, a discontinuous transition from a Mott localized state to a symmetry breaking polaronic state is observed, where double occupancy takes place in the two central dots. 
	The sharp transition allows for a continuous evolution when the hopping terms are finite, with setting in of pairing and phonon correlations.
	The transition can be detected in the zero hopping limit by measuring the occupation of the different dots with nearby single-electron transistors \cite{Hensgens2017} (Fig.~\ref{fig1}).
	The obtained results indicate that the system is particularly rich and is interesting to investigate both experimentally and theoretically because of the correlated states generated by the interplay of the electrons and phonons degrees of freedom.

	\begin{figure*}
		\centering
		\includegraphics[width=2.0\columnwidth]{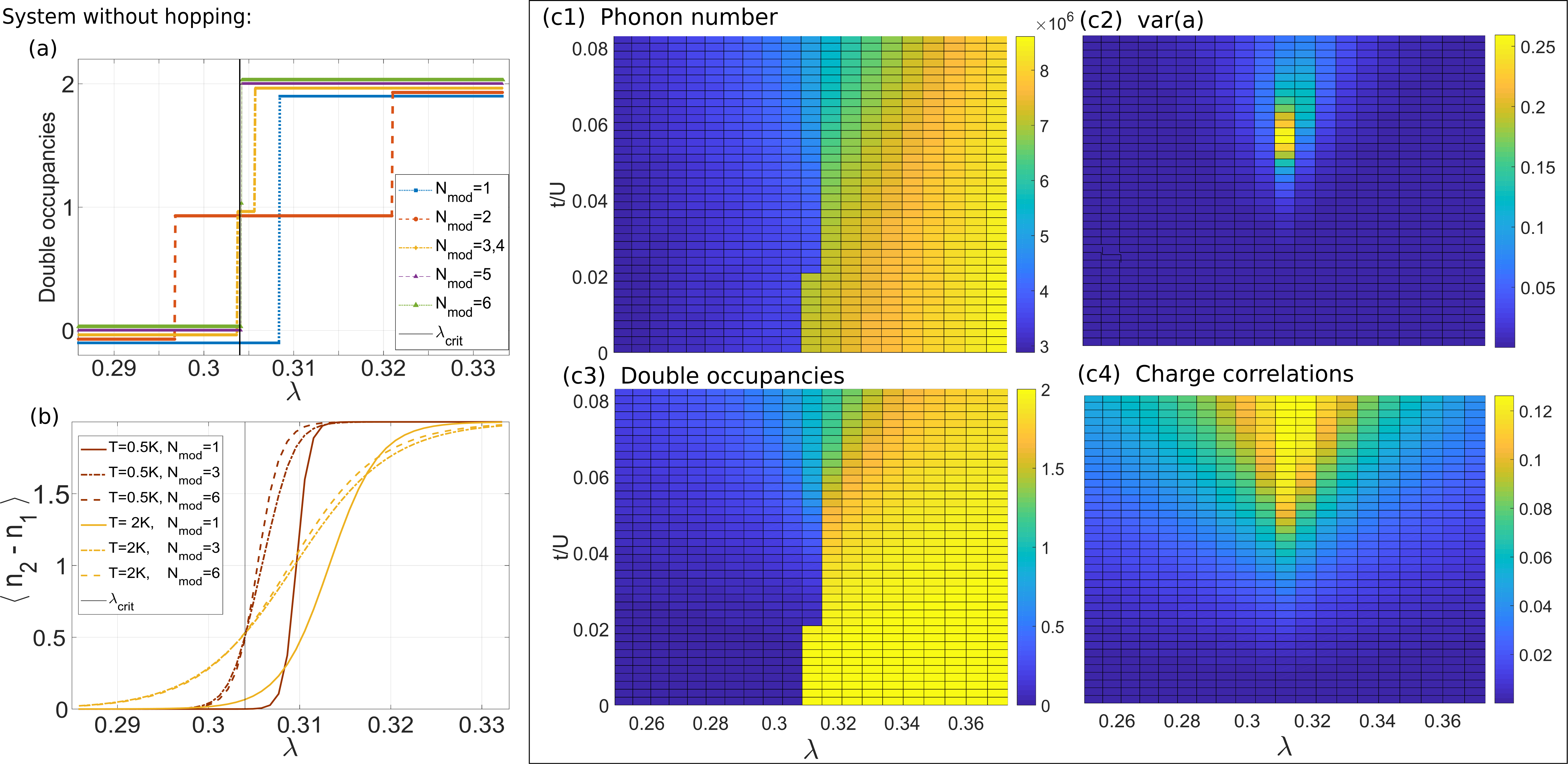}
		\caption{ (a,b) System behavior at $t=0$: As a function of the dimensionless coupling parameter $\lambda = g_0^2/\omega U$, we plot the number of doubly occupied sites in the ground state (a), and the density contrast between dot 1 and dot 2, $\langle n_2-n_1 \rangle$, at finite temperature (b).  The extent of the intermediate regime (1 double occupancy), between the Mott state (no double occupancy) and the paired state (2 double occupancies), depends on the number of modes $N_{\rm mod}$ taken into consideration. The density contrast between these opposite regimes is clearly visible even at $T=2K$. (c) System behavior at finite hopping: As a function $\lambda$ and $t$, we plot (c1) the number of phonons, (c2) the variance of the phonon operator, (c3) the number of double occupancies, (c4) the average value of charge correlations in the ground state. We consider a system of four electrons in four dots, with $U/(2\pi\hbar)=2400$GHz, and $\omega/(2\pi) = 1.35$MHz. In (c), we have restricted ourselves to a single-mode model, evaluated by exact diagonalization in a Hilbert space of up to 30 phonons, by iteratively shifting the phonon vacuum towards the correct ground state value. Qualitatively, the parameter space at small $t/U$ is divided into two localized regimes (Mott and paired regime), while at sufficiently large $t/U$ a delocalized regime occurs, as indicated by finite charge correlations.
		}
		\label{fig}
	\end{figure*}
	%
	%
	We consider a suspended carbon nanotube with four equally spaced quantum dots (see Fig.~\ref{fig1}). 
	We assume a perfectly symmetric device, where the three hopping terms $t$ between the four dots and local chemical potentials are equal. 
	In this configuration the local Coulomb repulsion $U$ is considered to be the same in each dot. 
	We assume that the interdot Coulomb coupling is negligible. 
	The system is prepared with 
	only four electrons populating the dots.
	The tunnelling amplitude to the leads of the first and fourth dots is assumed to be negligible, so that the total number of electrons remain fixed. 
	The charge on the dots naturally couples to the flexural modes of the carbon nanotube (see for instance \cite{pistolesi_proposal_2020}). 
	The Hamiltonian describing the system belongs to the EPCM  and therefore has a form:
	$H=H_{\rm e}+H_{\rm p}+H_{\rm e-p}$. 
	The electronic part is $H_{\rm e}= H_t + H_U$ with 
	$H_t = -t \sum_{i,\sigma} c_{i+1,\sigma}^\dagger c_{i,\sigma} + {\rm h.c.}$ 
	and a Hubbard term
	$H_U= \frac{U}{2} \sum_i n_i(n_i-1)$, 
	where $n_i=\sum_\sigma c_{i,\sigma}^\dagger c_{i,\sigma}$. 
	Here, the index $i$ represents the quantum dots, $\sigma=\pm$ accounts for the electrons' internal degree of freedom, and 
	$c_{i,\sigma}$ are the destruction operators for the electronic states.
	We will focus on the case of two degrees of freedom, corresponding either to the valley degrees of freedom for the spin-polarized electrons, or the spin degrees of freedom when the valley symmetry is broken. 
	The parameters $t$ and $U$ set the energy scale of hopping and on-site interaction. 
	Since the system is isolated from the leads we can set the chemical potential to zero.
	The phonon part reads $H_{\rm p} = \sum_\mu \hbar \omega_\mu a_\mu^\dagger a_\mu$. 
	Here $a_\mu$ is the destruction operator for the flexural mode $\mu$.
	We assume the limit of strong tension (guitar string limit) for which
	the resonating angular frequency $\omega_\mu= \mu \omega_0$ of the different modes is an integer multiple of the fundamental mode 
	frequency $\omega_0$.
	The e-p coupling reads $H_{\rm e-p}=\sum_{i,\mu} g_{i,\mu} n_i (a_\mu + a_\mu^\dagger)$, 
	with its strength set by a electrostatically tunable parameter $g_0$ \cite{micchi_mechanical_2015}, and explicitly  
	given by $g_{i,\mu}=g_0\frac{8}{\pi}\mu^{-3/2}\sin[\pi \mu(2i-1)/8 ] \sin[\pi\mu/8]$.
	This expression is obtained by expanding the functional dependence of the capacitance, between each dot and the gate, on the displacement of the nanotube, and integrating it for the total dot extension, which is assumed to be 1/4 of the total nanotube length.
	The hopping term $t/(2\pi\hbar)$ and the electron phonon coupling $g/(2\pi\hbar)$ can be electrostatically tuned between 1 and 100 GHz and between 0.01 and 1 GHz, respectively. The other parameters can be controlled by fabrication. Typically, the repulsion is $U/(2\pi\hbar) \sim $ 2 THz, while the fundamental mode $\omega_0/(2\pi)$ can range between 1 MHz and 1 GHz.
	(here $\hbar$ is the reduced Planck constant).
	
	%
	%

	Formally, the e-p coupling can be removed from the Hamiltonian by a Lang-Firsov (LF) transformation, $U=e^S$ with $S=\sum_{i,\mu} \frac{g_{i,\mu}}{\omega_\mu} n_i (a_\mu^\dagger -a_\mu)$.
	This transformation accounts for the phonon coupling through an additional effective interaction between electrons:
	\begin{align}
		H_{\tilde U} = - \sum_\mu \sum_{i,j} \frac{g_{i,\mu}g_{j,\mu}}{\omega_\mu} n_i n_j,
	\end{align}
	which is long-range due to the non-local nature of the phonons, and which can induce electron attraction. 
	In addition to generating this effective potential, the LF transformation also modifies the effective hopping, making it prone to numerical instabilities.
	Therefore, we use the LF transformation only in the atomic limit, $t=0$, where arbitrary numerical accuracy is possible.
	In this limit, the ground state is described by an electronic Fock state, which is selected by the potential $H_U + H_{\tilde U}$, times an effective phononic vacuum.
	The dimensionless parameter $\lambda=g_0^2/(\omega U)$ controls whether the Hubbard repulsion $H_U$ selects the Mott-insulating configuration with one electron per dot (i.e. the ground state expectation value $\langle n_i \rangle_0 =1$ for all $i$), or whether the phonon-induced interaction $H_{\tilde U}$, dominated by the lowest-energy mode $\mu=1$, depletes the system on the outer dots and induces electron pairing on the inner dots. 
	For concreteness, let us consider a nanotube with $N=4$ quantum dots, at half-filling and unpolarized with respect to $\sigma$, i.e. we have $N/2$ electrons with $\sigma=+$, and $N/2$ electrons with $\sigma=-$. 
	We find that the $N!/[(N/2)!]^2=6$ Mott states (i.e. all states with occupation numbers $\{1,1,1,1\})$ have energy $E/U=-\frac{2\pi^2}{3} \lambda$, and the unique paired state with occupation numbers $\{0,2,2,0\}$ has $E/U=2-\frac{4\pi^2}{3}\lambda$.
	Accordingly, these two sets of states provide the electronic ground states of the system  for any value of $\lambda$, with a level crossing at $\lambda_{\rm crit}= 3/\pi^2$. 
	We note that in this critical point, there are four additional ground states, referred to as ``intermediate'' states and characterized through occupation numbers $\{1,2,1,0\}$ or $\{0,1,2,1\}$ with an energy-dependence $E/U=1-\pi^2\lambda$. 
	To obtain these expressions for the energies, we have carried out an infinite sum over the phonon modes, although for practical purposes one might instead truncate this sum at a finite number of modes $N_{\rm mod}$. 
	In this case, the scenario in the critical region is slightly altered: The degeneracy of three different types of density patterns (Mott, paired, intermediate) is lost. Instead, the intermediate  states may appear above the ground states for any  $\lambda$. This happens, e.g., for $N_{\rm mod}=1$ or $N_{\rm mod}=5$.  Preferentially, though, the truncation of the modes gives rise to a tiny but finite parameter regime, in which the intermediate states become the unique ground states. This behavior is illustrated in Fig.~\ref{fig}(a), where the three regimes are distinguished by the number of doubly occupied sites in the ground state, plotted versus the {\color{blue}} dimensionless coupling parameter $\lambda$.
	
	In practice, the number of doubly occupied dots is hard to quantify, since it would require simultaneous measurements of all local densities. However, as shown in Fig.~\ref{fig}(b), the density estimation on only two sites is sufficient to clearly distinguish between the different regimes. Concretely, we evaluate the density difference between an inner and an outer dot $\langle n_2-n_1 \rangle$. This quantity takes the value 2 in the paired state, 1 in the intermediate state, and 0 in the Mott state. Here, instead of ground state averages, we have considered thermal averages at temperatures between 0.5K and 2K, assuming that the system's energy scale is given by an interaction parameter $U/(2\pi\hbar)=2400$GHz. Moreover, we have varied the number of modes, $N_{\rm mod}=$ 1, 3, or 6. Importantly, in all cases and for all temperatures, the transition from Mott to paired state is clearly seen from this data, with a steeper change at $T=0.5$K. At larger $T$, the intermediate regime broadens, and notably, the broadening is more pronounced in the 3- or 6-mode model than in the one-mode model, where the intermediate state is absent from the ground state manifold. In none of these cases, though, the intermediate state would give rise to a flat regime at $\langle n_2-n_1 \rangle=1$, demonstrating the secondary role played by the intermediate states.
	
	
	The LF transformation can be viewed as a polaron dressing of the electrons, in which  a certain electron occupation $n_i$ implies the presence of $N_\mu=\langle a_\mu^\dagger a_\mu \rangle = \left( \sum_i n_i  \frac{g_{i,\mu}}{\omega_\mu} \right)^2$ phonons in mode $\mu$. From this number, readily obtained at $t=0$, we find valuable information also for the system at small but finite $t$. As mentioned above, numerical instability then prohibit the use of the LF transformation, and truncation of the phononic Hilbert space becomes necessary. According to the expression for $N_\mu$, we find that, near criticality, the first (and most occupied) mode has $N_1 \approx (8\lambda_{\rm crit} U/g_0)^2$ phonons. Thus, for the experimentally most relevant case of $U\gg g_0$, it is impossible to treat a Hilbert space with ${\cal O}(N_1)$ phonons. 
	
	Instead, we have developed an iterative shift method based on making the replacement
	\begin{align}
		a_\mu  \rightarrow \tilde a_\mu + S_\mu,
	\end{align}
	where $\tilde a_\mu$ is a shifted phonon operator, and $S_\mu$ is a complex number. If we choose $S_\mu = \sqrt{N_\mu}$, and numerically diagonalize the full Hamiltonian within a highly truncated Hilbert space of tilded phonons (e.g., $n_{\rm max} \approx 30$), we recover, at $t=0$, the same result as obtained before using the LF transformation. At finite $t$, the shift parameter $S_\mu$ has to be adjusted properly, which can be done iteratively: Using the $t=0$ shift as an initial guess, we determine a new shift parameter 
	\begin{align}
		S_\mu' = \sqrt{\langle (\tilde a_\mu^\dagger + S_\mu^*)(\tilde a_\mu +S_\mu) \rangle_0},
	\end{align}
	and repeat updating this parameter, until $S_\mu$ and $S_\mu'$ agree with the desired numerical precision. We have verified this method at small values of $U/g_0 \sim 1$, where a numerical procedure without shift is also possible, due to the relatively small number of phonons. 
	
	In the following, we report on our results for the experimentally realistic values $U/(2\pi\hbar)=2400$GHz, being much larger than the phonon coupling strength $g_0$ (tunable, on the order of 1GHz \cite{vigneau_ultrastrong_2021}), and the phonon frequency, set to $\omega/(2\pi)=1.35$MHz in order to exploit the system near criticality. Such low resonance frequencies can be achieved in long nanotubes \cite{Moser2013}. In this scenario, the phonon number is ${\cal O}(10^6)$, and therefore, we fully rely on the shift method.

	Our results are shown in Fig.~\ref{fig}(c1--c4) for a one-mode model. Qualitatively, we find three regimes: At small $t/U$, there are the two localized regimes (Mott and paired regime), clearly distinguished by the occupation of the dots [e.g. number of double occupancies plotted in Fig.~\ref{fig}(c3)], but also through an abrupt change of the number of phonons [Fig.~\ref{fig}(c1)]. At sufficiently large $t/U$, a delocalized regime occurs, characterized through an intermediate phonon number and an intermediate number of doubly occupied dots. Interesting features of this third regime are finite values of electronic correlations, and the correlated
	phonon state. The latter is indicated by non-vanishing values of ${\rm var}(a) \equiv \langle a^\dagger a \rangle -\langle a^\dagger \rangle \langle a\rangle$ [Fig.~\ref{fig}(c2)]. As an illustration of the electronic correlations, we plot in Fig.~\ref{fig}(c4)  the average charge correlations $C$ given by
	\begin{align}
		C = \frac{1}{N} \sum_{i,j} \left( \langle n_i n_j \rangle_0 - \langle n_i\rangle_0 \langle n_j \rangle_0 \right).
	\end{align}
	We note that in this regime, we have also obtained finite values of other pair correlators, such as s-wave or p-wave pairing correlators, $\langle S_i S_j \rangle_0$ or $\langle P_i P_j \rangle_0$, with $S_i^\dagger = c_{i\uparrow}^\dagger c_{i\downarrow}^\dagger$, and $P_i^\dagger = (c_{i+1\uparrow}^\dagger c_{i\downarrow}^\dagger + c_{i+1\downarrow}^\dagger c_{i\uparrow}^\dagger)/\sqrt{2}$. 
	
	The correlated nature makes the delocalized state at large $t$ crucially different from the intermediate state at $t=0$ discussed earlier, despite their similar structure of the density. We stress that the delocalized regime is \textit{not} adiabatically connected to the intermediate $t=0$ states, as we have checked (at a small value of $U$) by a three-mode calculation, which explicitly exhibits $t=0$ intermediate states [cf. Fig.~\ref{fig1}(a)]. Specifically, this calculation has shown that (i) finite values of $t$ suppress the intermediate state until it fully vanishes, and that (ii) the delocalized regime, characterized by finite pair correlations, occurs only at even larger values of $t$. Intuitively, the suppression of the intermediate state by the hopping is understood from the extremely limited amount of first-order hopping which are possible in this configuration.
	
	In summmary, we have proposed an experimentally feasible way for quantum simulation of a Hamiltonian belonging to the EPCM, by placing four equally placed quantum dots in a suspended nanotube. The presence of e-e interactions and e-p interactions leads to a competition between different types of states: a Mott state dominated by the e-e interaction, a polaronic state dominated by the e-p interaction, and a strongly correlated delocalized state at large hopping. At small hopping, we also find an intermediate state, but it is not adiabatically connected to the strongly delocalized state. We have theoretically explored the system by employing the Lang-Firsov transformation, which gives us analytically exact results for zero hopping, and by developing a numerically self-consistent iterative shift method. The distinction between the different regimes is done by looking at quantities such as the local electron density and e-e correlators, or the phonon distribution. These quantities in the finite hopping limit may be accessible in experiments where two of the dots are locally coupled to two different superconducting resonators \cite{viennot_coherent_2015}. Moreover, in the proposed experimental setup, the parameters are sufficiently tunable to explore the different regimes, and we find that the stability of the states against finite temperatures is well within the feasible temperature bounds.

	Although this work focuses on the quantum simulation of the competition between the e-e and e-p coupling in four quantum dots embedded in a nanotube, it should be noted that our overall scheme is quiet general.
	The system allows for the investigation of many intriguing scenarios. Those may arise when controllable quantities like the number of electrons or dots, and the 
	nature of mechanical modes (guitar string or doubly clamped beam without tensile tension) are varied.
	When filling $N/2$ electrons in $N$ dots, the Peierls transition and charge density wave states may emerge with the help of the inter-dot Coulomb repulsion to prevent electrons to localize in nearby dots.
	The internal degree of freedom (spin and/or valleys) can also lead to new kinds of order, which can become relevant for larger number of dots.
	Finally, by considering a current flowing through the nanotube, non-equilibrium physics could also be explored in the presence of e-p coupling, in such setups.

	\begin{acknowledgements}
		ICFO acknowledges support by
		Severo Ochoa (grant number SEV-2015-0522),
		Fundació Cellex, Fundació Mir-Puig,
		the CERCA Programme, 
		the Fondo Europeo de Desarrollo Regional,
		European Social Fund.
		A.B. acknowledges support by 
		ERC advanced (grant number 692876), 
		ERC PoC (grant number 862149), 
		AGAUR (grant number 2017SGR1664),  
		MICINN (grant number RTI2018-097953-B-I00).
		M.L., U.B., and T.G. acknowledge support by 
		ERC AdG NOQIA, 
		Spanish Ministry MINECO and State Research Agency AEI (FIDEUA PID2019-106901GB-I00/10.13039 / 501100011033, and CEX2019-000910-S, FPI),   Generalitat de Catalunya (AGAUR Grant No. 2017 SGR 1341, QuantumCAT U16-011424, co-funded by ERDF Operational Program of Catalonia 2014-2020), 
		MINECO-EU QUANTERA MAQS (funded by State Research Agency (AEI) PCI2019-111828-2 / 10.13039/501100011033), 
		EU Horizon 2020 FET-OPEN OPTOLogic (Grant No 899794), and the National Science Centre, 
		Poland-Symfonia Grant No. 2016/20/W/ST4/00314. 
		T.G. acknowledges funding from ``la Caixa'' Foundation (ID 100010434, fellowship code LCF/BQ/PI19/11690013). 
		U.B. acknowledges support by the “Cellex-ICFO-MPQ Research Fellows", a joint program between ICFO and MPQ – Max-Planck-Institute for Quantum Optics, funded by the Fundació Cellex.
		F.P. acknowledges support from the French Agence Nationale de la Recherche (grant SINPHOCOM ANR-19-CE47-0012) and Idex Bordeaux (grant Maesim Risky project 2019 of the LAPHIA Program).
	\end{acknowledgements}
	
	\bibliography{FabioBiB.bib}
\end{document}